\documentclass[twoside]{article}
\usepackage{amsmath,amssymb}
\usepackage{qic}

\def\>{\ensuremath{\rangle}}
\def\<{\ensuremath{\langle}}

\def\-{\ensuremath{\textrm{-}}}

\def\h{\ensuremath{\mathcal{H}}}

\def\f{\ensuremath{\mathcal{F}}}
\def\m{\ensuremath{\mathcal{M}}}

\def\k{\ensuremath{\mathcal{K}}}

\def\e{\ensuremath{\mathcal{E}}}
\def\f{\ensuremath{\mathcal{F}}}

\def\<{\langle}
\def\>{\rangle}

\def\k{\mathcal{K}}

\newtheorem{example}{Example}

\begin{document}
\setlength{\textheight}{8.0truein}    

\runninghead{Super-activating quantum memory with entanglement}
            {J. Guan, Y. Feng, and M-S Ying}

\normalsize\textlineskip
\thispagestyle{empty}
\setcounter{page}{1115}

\copyrightheading{18}{13\&14}{2018}{1115--1124}

\vspace*{0.88truein}

\alphfootnote

\fpage{1115}

\centerline{\bf
SUPER-ACTIVATING QUANTUM MEMORY WITH ENTANGLEMENT}

\vspace*{0.37truein}

\centerline{\footnotesize 
JI GUAN\footnote{Corresponding author. Email: guanji1992@gmail.com.}}
\vspace*{0.015truein}

\centerline{\footnotesize\it State Key Laboratory of Computer Science, Institute of Software}
\baselineskip=10pt
\centerline{\footnotesize\it  Chinese Academy of Sciences, Beijing 100190, China}
\baselineskip=15pt
\centerline{\footnotesize\it Center for Quantum Software and Information,
University of Technology Sydney, NSW 2007, Australia}
\vspace*{10pt}
\centerline{\footnotesize
YUAN FENG 
}
\vspace*{0.015truein}
\centerline{\footnotesize\it Center for Quantum Software and Information,
University of Technology Sydney, NSW 2007, Australia}
\baselineskip=10pt
\vspace*{10pt}
\centerline{\footnotesize 
MINGSHENG YING}
\vspace*{0.015truein}
\centerline{\footnotesize\it Center for Quantum Software and Information,
University of Technology Sydney, NSW 2007, Australia}
\baselineskip=15pt
\centerline{\footnotesize\it State Key Laboratory of Computer Science, Institute of Software}
\baselineskip=10pt
\centerline{\footnotesize\it  Chinese Academy of Sciences, Beijing 100190, China}
\baselineskip=15pt
\centerline{\footnotesize\it Department of Computer Science and Technology, Tsinghua University, Beijing 100084, China}
\vspace*{0.225truein}
\publisher{June 2, 2018}{September 30, 2018}

\vspace*{0.21truein}

\abstracts{
Noiseless subsystems were proved to be an efficient and faithful approach to preserve fragile information against decoherence in quantum information processing and quantum computation. They were employed to design a general (hybrid) quantum memory cell model that can store both quantum and classical information. In this paper, we find an interesting new phenomenon that the purely classical memory cell can be super-activated to preserve quantum states, whereas the null memory cell can only be super-activated to encode classical information. Furthermore, necessary and sufficient conditions for this phenomenon are discovered so that the super-activation can be easily checked by examining certain eigenvalues of the quantum memory cell without computing the noiseless subsystems explicitly. In particular, it is found that entangled and separable stationary states are responsible for the super-activation of storing quantum and classical information, respectively.}{}{}

\vspace*{10pt}

\vspace*{3pt}
\communicate{R~Jozsa~\&~A~Harrow}

\vspace*{1pt}\textlineskip    

\section{Introduction}
\noindent
Storing, transmitting and transforming information are the basic tasks in quantum information processing. It is crucial to develop techniques to reduce and correct errors when these tasks are being processed. A passive way for this purpose is to encode information into certain \emph{noiseless subsystems} which can overcome quantum noises~\cite{lidar2012review}. However, we can also use an active method that transforms a noisy subsystem to be noiseless by performing some correction operation. This method is termed as \emph{operator quantum error correction}~\cite{kribs2005unified,kribs2005operator}. These two approaches are usually complementary to each other in avoiding quantum errors \cite{knill2006protected}. Recently, two general frameworks using both the passive and active methods for perfectly protecting information have been developed, namely \emph{information-preserving structures} \cite{blume2008characterizing} and \emph{operator algebra quantum error correction} \cite{beny2007generalization}. 
In particular, these approaches provide efficient ways for the protection of hybrid quantum-classical information.

Quantum memory is often considered as a method of delayed usage of quantum states and a set of quantum memory cells.
The general \emph{(hybrid) quantum memory cell} was first introduced in \cite{kuperberg2003capacity} and was  also termed as \emph{noiseless information-preserving structure} \cite{blume2010information}. The memory cell $\m$ has separate (orthogonal) sectors, and each sector is a noiseless subsystem, labeled by a classical ``address''. Quantum information is stored in each sector with the dimension being at least two,  whereas classical information is encoded by the choice between the different sectors. 
It is reasonable to assume that in a quantum memory,  the noise $\e$ (characterized as a quantum operation) on every memory cell is independent of each other and can happen over and over again, while the number of the occurrence is unknown but the same on every memory cell.  Moreover, the memory cell is uniquely determined by the noise $\e$ \cite{blume2010information}, so we can define memory cell $\m$ by quantum operation $\e$. For simplicity, we directly use $\e$ for $\m$. 
However, useful noiseless subsystems (quantum information can be stored) are limited to the noise that contains some symmetries. Such symmetries are often absent in real devices \cite{wang2016minimal,wang2013numerical}, and thus the memory cell can only preserve classical information if there are at least two sectors; in other words, it is degenerated to be \emph{purely classical}.
 
 In this paper, we consider two strategies of using quantum memory cells: individually and collectively.
 In particular, we observe an interesting new phenomenon: a purely classical memory cell $\e$  can be super-activated to store quantum information; that is, collective use of memory cells can perfectly preserve quantum states while individual memory cell cannot. We find that the memory cell $\e$ can be super-activated if and only if there is an entangled stationary state of  $\e^{\otimes 2}$. Furthermore, we can give a simple characterization of the super-activation in terms of \emph{external} eigenvalues (with magnitude one) of $\e$. This enables us to easily check the super-activation property as well as the existence of entangled stationary states.  Once activated, the maximum dimension of sectors in the memory cell $\e$ may have an exponential growth with the number of used memory cells. Moreover, we see that a \emph{null memory cell} (no information can be protected) cannot be super-activated to store quantum information, but classical information can be preserved with the collective use of it. This property can be characterized by \emph{internal} eigenvalues. 

{\vskip 4pt}

\section{Quantum memory}
\noindent
Given a quantum system $S$ with the associated (finite-dimensional) state space $\h$. We say that a quantum system $A$ is a \emph{subsystem} of $S$ if  
$\h=(\h_A\otimes \h_B)\oplus\k$ 
for some quantum system $B$, where $\h_A$ and $\h_B$ are the state spaces of $A$ and $B$, respectively. Furthermore, if dim$(\h_B)=1$, then $\h_A$ is a \emph{subspace}. 
Let $B(\h)$ be the set of all linear operators on $\h$ and $D(\h)$ the set of \emph{quantum states}, i.e. density operators with unit trace, on $\h$.
The \emph{support}  of a quantum state $\rho$, denoted by supp$(\rho)$, is the linear span of the eigenvectors corresponding to non-zero eigenvalues of $\rho.$ A \emph{quantum operation} on $\h$ can be represented by a \emph{completely positive and trace-preserving} (CPTP) map from operators on $\h$ to themselves. Quantum noises and quantum channels both are concrete instances of quantum operations.

\begin{definition}
        Given a quantum operation $\e$ on $\h=(\h_A\otimes \h_B)\oplus\k$, subsystem $A$ is said to be \emph{noiseless} if
         for any $\rho_A$ and $\rho_B$, there exists a quantum state $\xi_B$ such that $\e(\rho_A\otimes \rho_B)=\rho_{A}\otimes \xi_{B}.$
\end{definition}

From the above definition, we see that $\h_A$ is kept intact by the operation $\e$.
It was shown in \cite{choi2006method} that noiseless subsystems of $\e$ can be characterized by the set of its \emph{fixed points}, denoted by $fix(\e)=\{A\in B(\h)\mid \e(A)=A\}$. If $\rho\in fix(\e)$ is a quantum state, then we call it a \emph{stationary state}; furthermore, if there is no other stationary state $\sigma$ with  supp$(\sigma)\subseteq \textrm{supp}(\rho)$, then $\rho$ is said to be \emph{minimal}. From \cite{blume2010information,wolf2012quantum}, with an appropriate decomposition of the Hilbert space $\h=\bigoplus_{k=1}^n(\h_{A_k}\otimes \h_{B_k})\oplus \mathcal{K}$, the fixed points admit a useful structure: 
\begin{eqnarray}\label{eq_decomposition}
    fix(\e)=\bigoplus_{k=1}^{n}\left(B(\h_{A_k}) \otimes \sigma_k\right) \oplus 0,
\end{eqnarray}
where $\sigma_k$ is  a full-rank quantum state on $\h_{B_k}$ and $\h_{A_i}\otimes \h_{B_i}$ is orthogonal to $\h_{A_j}\otimes \h_{B_j}$ if $i\not=j$.   This decomposition is unique (up to the order of $k$) and called \emph{the fixed-point decomposition} of $\e$ and can be computed by the algorithms in \cite{knill2006protected,guan2018decomposition}. It is easy to see that for each $k$, $\h_{A_k}$ is a noiseless subsystem. Conversely, this decomposition captures all noiseless subsystems; that is, $\h_A$ is a noiseless subsystem if and only if $\h_A\subseteq \h_{A_k}$ for some $k$. 


Using the fixed-point decomposition of a given quantum operation $\e$, we can partition $\eta(\e)$, its multiset of eigenvalues with magnitude one, into two parts. For each $k$, let $\eta_k(\e)$ be the multiset of  \emph{internal} eigenvalues of $\e$ restricted on $\h_{A_k}\otimes \h_{B_k}$ in Eq.(\ref{eq_decomposition}), again with magnitude one, and
let $\bar{\eta}(\e)=\eta(\e)\setminus \cup_k\eta_{k}(\e)$ be the  \emph{external} eigenvalues of $\e$. 

To perfectly protect quantum information under a quantum noise $\e$, a quantum memory cell $\m$ is defined in \cite{kuperberg2003capacity,blume2010information} as the structure of noiseless subsystems in Eq.(\ref{eq_decomposition}) of $\e$ to store quantum states into each noiseless quantum subsystem $A_k$  if the dimension $d_k>1$ of $\h_{A_k}$. $\m$ can preserve quantum states against any power of $\e$, even though the exact number of  $\e$ happening is unknown. Therefore, $\m$ is useful in practice, and this is also the reason why we choose noiseless subsystems as the basis of quantum memory cells. On the other hand, if we can record the number of applications of $\e$, then quantum memory cells can be designed on more general reversible subsystems, called \emph{decoherence-free subsystems} \cite{shabani2005theory}.

\begin{definition}
Given a quantum operation $\e$ on $\h=(\h_A\otimes \h_B)\oplus\k$, subsystem $A$ is said to be \emph{decoherence-free} if we can find a unitary matrix $U_A$ on $\h_A$ such that 
         for any $\rho_A$ and $\rho_B$, there exists a quantum state $\xi_B$ such that $\e(\rho_A\otimes \rho_B)=U_A\rho_{A}U_A^\dagger\otimes \xi_{B}.$     
 \end{definition} 

 Noiseless subsystems are a special case of decoherence-free subsystems. As $U_A$ is reversible, decoherence-free subsystems can protect quantum information from any power of $\e$ when the number of the power is kept.

Through the above discussions, $\m$ is characterized by $\e$, so in the following, for simplicity, we directly use $\e$ for $\m$.
Define the \emph{shape} of memory cell $\e$ as $\lambda(\e)=(d_{1},\cdots,d_{n})$.  As the structure of $fix(\e)$ is unique (up to the order of $k$), $\lambda(\e)$ is well-defined. In a sense, $\lambda(\e)$ represents the capacity of the memory cell $\e$; that is, how much quantum and classical information can be preserved: \begin{itemize}\item $|\lambda(\e)|_{\infty}=\max_{k}d_k$ is the largest dimension of quantum states that can be stored; 
\item the length $|\lambda(\e)|=n$  is the preserved classical information as the choice between the different sectors.\end{itemize} If $|\lambda(\e)|_{\infty}=1$ and $|\lambda(\e)|>1$, then the quantum memory cell is degenerated to be purely classical, and quantum information cannot be preserved. Furthermore, the memory cell is said to be \emph{null} if $\lambda(\e)=(1)$; that is, neither quantum nor classical information can be stored.  

 The quantum memory cell model encompasses the existing techniques for preserving quantum and classical information: noiseless subsystems, pointer basis \cite{zurek1981pointer} and decoherence-free subspaces \cite{lidar2012review}. 
For example, pointer bases have
the shape $(1,1,...,1)$, describing a complete set of one-dimensional $k$ sectors (both $A_k$ and $B_k$ are trivial for all $k$).
A decoherence-free subspace has the shape $(d)$, describing a single $k$ sector with a trivial $B_k$. 

In the following sections, we will concern the \emph{super-activations} of quantum memory cells.

\begin{definition}
     Given a quantum memory cell $\e$, 
     \begin{itemize}
          \item $\e$ is said to be  super-activated to preserve quantum information, if $\e$ cannot store quantum states but $\e^{\otimes 2}$ can, i.e.,
          $|\lambda(\e)|_\infty=1$ and $|\lambda(\e^{\otimes2})|_\infty>1$; for simplicity, we can directly say that $|\lambda(\e)|_{\infty}$ is super-activated. 
          \item $\e$ is said to be  super-activated to preserve classical information, if $\e$ cannot store classical bits but $\e^{\otimes 2}$ can, i.e.,  $|\lambda(\e)|=1$ and $|\lambda(\e^{\otimes2})|>1$; for simplicity, we can directly say that $|\lambda(\e)|$ is super-activated. 
      \end{itemize} 
 \end{definition}
 
{\vskip 4pt}

\section{Super-activation for storing classical information} 
\noindent
Given a quantum memory cell $\e$, if its shape is $(1,\cdots,1)$, then quantum information cannot be stored in it. Indeed, it was shown in \cite{wang2016minimal,wang2013numerical} that in practice only a very small set of quantum memory cells admits a useful noiseless subsystem (with the dimension being at least 2). Fortunately, this problem can be remedied by the collective use of quantum memory cells where a super-activation of $|\lambda(\cdot)|_{\infty}$ can happen. 
Let us start with the simplest case with the shape $\lambda(\e)=(1)$, i.e., a null memory cell. Typical examples include irreducible quantum channels (having only one stationary state with the full rank) and amplitude damping channels. In this case,  memory cell $\e$ behaves periodically in some subspaces.
\begin{lemma}\label{lem_eigen}
    Let $\e$ be a quantum memory cell with $\lambda(\e)=(1)$. Then there exists some integer $p$ such that 
\begin{enumerate}
    \item $\eta(\e)=\{\exp(2\pi ik/p)\}_{k=0}^{p-1}$ with each element being internal;
    \item a set of mutually orthogonal states $\{\rho_{i}\}_{i=0}^{p-1}$ can be found such that $\e(\rho_{i})=\rho_{i\boxplus 1}$, where  $\boxplus$ denotes addition modulo $p$.
\end{enumerate}    
\end{lemma}
{\it Proof.}
See the Appendix. 
\hfill $\Box$

The integer $p(\e)=p$ in the above lemma is called the \emph{period} of $\e$. Note that $p(\e)$ is the number of internal eigenvalues.
Apparently, the memory cell $\e$ cannot store quantum and classical information. However, we can create new shelters for classical information by using two quantum memory cells with the simplest shapes; that is, $|\lambda(\cdot)|$ can be activated for storing classical information. The following is a simple example of such super-activation.

\begin{example}
    Let $\e$ be  a quantum memory cell on $\h=\textrm{lin.span}\{|0\rangle, |1\rangle\}$ with 
    $\e(\cdot)=|0\rangle\langle 1 |\cdot|1\rangle\langle 0 |+|1\rangle\langle 0 |\cdot|0\rangle\langle 1 |.$ 
    It is easy to see that $\lambda(\e)=(1)$ and $\lambda(\e^{\otimes 2})=(1,1)$, indicating that $|\lambda(\e)|$ is activated.
\end{example}

A general characterization of super-activation for storing classical information is presented in the following: 

\begin{theorem}\label{theorem_one_shape}
    For any two quantum memory cells $\e$ and $\f$ with $|\lambda(\e)|=|\lambda(\f)|=1$,  there exists a set of mutually orthogonal quantum states $\{\rho_{i}\}_{i=0}^{m-1}$ with $m=gcd\{p(\e),p(\f)\}$, the greatest common divisor of $p(\e)$ and $p(\f)$,  such that:
    $$fix(\e \otimes\f)=\bigoplus_{i=0}^{m-1} \rho_{i}\oplus 0$$ that is, $\lambda(\e\otimes\f)=(1,\cdots,1)$ and $|\lambda(\e\otimes\f)|=m$. Furthermore, for each $i$, $\rho_{i}$ is separable.
\end{theorem}
{\it Proof.} As $|\lambda(\e)|=|\lambda(\f)|=1$, there is only one stationary state $\sigma$ and $\rho$ for $\e$ and $\f$, respectively. By Lemma~\ref{lem_eigen}(2), there are two sets of mutually orthogonal states $\{\sigma_{i}\}_{i=0}^{p(\e)-1}$ and $\{\rho_{i}\}_{i=0}^{p(\f)-1}$  such that $\e(\sigma_{i})=\sigma_{i\boxplus1}$ and $\f(\rho_{i})=\rho_{i\boxplus1}$.  For $0\leq i\leq m-1$, let 
\[
\sigma^{i}=\frac1{K_\e}\sum_{j=0}^{K_\e-1}\sigma_{i\boxplus jm},\ \ \ \ \ \rho^{i}=\frac1{K_\f}\sum_{j=0}^{K_\f-1}\rho_{i\boxplus jm}
\] 
where $K_\e = p(\e)/m$ and $K_\f = p(\f)/m$. We claim that $\{(\sum_{i}\sigma^{i}\otimes\rho^{i\boxplus j})/m\}_{j=0}^{m-1}$ is a set of mutually  orthogonal stationary state for $\e \otimes \f$. Indeed,  by Lemma \ref{lem_eigen}(1), we have $\eta(\e)=\{\exp(2\pi ik/p(\e))\}_{k=0}^{p(\e)-1}$ and  $\eta(\f)=\{\exp(2\pi ik/p(\f))\}_{k=0}^{p(\f)-1}$, so the multiplicity of eigenvalue one (for $\e\otimes\f$) is $m$. Thus for each $j$, $(\sum_{i=0}^{m-1}\sigma^{i}\otimes\rho^{i\boxplus j})/m$ is a minimal stationary state of $\e\otimes \f$. We finish the proof by noting that it has no other minimal stationary states.
\hfill $\Box$

The proof of Theorem \ref{theorem_one_shape} gives us an explicit way to construct the memory cell structure, namely the decomposition Eq.(\ref{eq_decomposition}). Usually, entanglement is responsible for the super-activation of many physical quantities in quantum information theory, such as the zero-error capacity of quantum channels \cite{cubitt2011superactivations}. However, the above theorem shows that it is not the case for $|\lambda(\cdot)|$.  

For the special case where multiple copies of $\e$ are collectively used, we have:  
\begin{corollary}
Suppose $\e$ is a null memory cell with $|\lambda(\e)|=1$. Then
\begin{enumerate}
\item    $|\lambda(\e^{\otimes t})|=p(\e)^{t-1}$ for any $t\geq 1$. That is, perfect storage of classical information can always be super-activated as long as $p(\e)>1$; 
\item $|\lambda(\e^{\otimes t})|_{\infty}=1$ for any $t\geq 1$. That is, no quantum information can be perfectly preserved even collective use of memory cells  is employed; 
\item All stationary states of $\e^{\otimes t}$ are separable.
\end{enumerate}
\end{corollary}
{\it Proof.} By the similar construction of stationary states in the proof of Theorem \ref{theorem_one_shape}, we obtain $p(\e)^{t-1}$ separable and mutually orthogonal stationary states for $\e^{\otimes t}.$ Then we compute $\lambda(\e^{\otimes t})=(1,\cdots, 1)$ and $|\lambda(\e^{\otimes t})|=p(\e)^{t-1}$ by noting that the multiplicity of eigenvalue one (for $\e^{\otimes t}$) is $p(\e)^{t-1}$. 
\hfill $\Box$

Note that the period of a null memory cell represents how much classical information can be activated. The above corollary shows that once activation happens, the amount of preserved information can grow up continuously with the number of the application of the memory cells.  

{\vskip 4pt}

\section{Super-activation for storing quantum information}
\noindent
  The results presented in the last section show that two null memory cells can be used together to super-activate the amount $|\lambda(\e)|$ of stored classical information. In this section, we are going to show how the amount $|\lambda(\e)|_{\infty}$ of stored quantum information can be super-activated. The following theorem gives a necessary and sufficient condition for this quantum super-activation. 

\begin{theorem}\label{Theo_ns_super}
    Let $\e$ and $\f$ be two quantum memory cells  with $|\lambda(\e)|_{\infty}=|\lambda(\f)|_{\infty}=1$. Then the following statements are equivalent:
    \begin{enumerate}
         \item [(1)]  $|\lambda(\e\otimes \f)|_{\infty}\geq 2$;
         \item [(2)] there exists an  entangled stationary state for $\e\otimes \f$;
         \item [(3)] there exist $a\in \eta(\e)$, $b\in \eta(\f)$ such that $ab=1$, and $a$ or $b$ is external, i.e.  $a\in \bar{\eta}(\e)$ or $b\in \bar{\eta}(\f)$. 
\end{enumerate}
\end{theorem}
{\it Proof.} The implications $(1)\Rightarrow (3)$ and $(2)\Rightarrow (3)$ are  from Theorem \ref{theorem_one_shape}. As $|\lambda(\e)|_{\infty}=|\lambda(\f)|_{\infty}=1$, $\lambda(\e)=(1,\cdots,1)$ and $\lambda(\f)=(1,\cdots,1)$; that is there are only finitely many mutually orthogonal minimal stationary states  $\{\rho_{i}\}$ and $\{\sigma_{j}\}$ for $\e$ and $\f$, respectively. If for any $a\in \eta(\e)$ and $b\in \eta(\f)$, $ab=1$ can only occur when $a$ and $b$ both are internal, i.e.  $a\not\in \bar{\eta}(\e)$ and $b\not\in \bar{\eta}(\f)$, then it is enough to restrict $\e$ and $\f$ onto the subspaces supp$(\rho_i)$ and supp$(\sigma_j)$ respectively, for each $i$ and $j$, when we compute $fix(\e\otimes \f).$ Furthermore, as for each $\rho_i$  and $\sigma_j$ are minimal, the restricted memory cells $\e_i$ and $\f_j$ have the simplest shapes. Therefore, following Theorem \ref{theorem_one_shape}, all stationary states are separable and $|\lambda(\e\otimes \f)|_{\infty}=1$, contradicting the assumptions $(1)$ or $(2)$.

To prove $(3)\Rightarrow (1)$ and $(3)\Rightarrow (2)$, let $A$ and $B$ be eigenvectors of $\e$ and $\f$ corresponding to eigenvalues $a$ and $b$ respectively, i.e. $\e(A)=aA$ and $\f(B)=bB$.
In the following, we only prove the case that $a\in \bar{\eta}(\e)$ and $b\in \bar{\eta}(\f)$. Other cases are similar. 

From the decomposition Eq.(\ref{eq_decomposition}) of $\e$ and $\f$, there exist mutually orthogonal minimal stationary states $\rho_1$, $\rho_2$ for $\e$ and  $\sigma_1$, $\sigma_2$ for $\f$ such that 
\begin{align*}
    A & \in \textrm{lin.span}\{|\psi_1\rangle\langle \psi_2|: |\psi_i\rangle\in \textrm{supp}(\rho_i), i=1,2\},\\
    B& \in \textrm{lin.span}\{|\phi_1\rangle\langle \phi_2|: |\phi_i\rangle\in \textrm{supp}(\sigma_i), i=1,2\}.
\end{align*}
Then we can find a positive number $\epsilon$ such that 
\[
\frac1K \left[\rho_{1}\otimes \sigma_{1}+\rho_{2}\otimes \sigma_{2}+\epsilon(A\otimes B+A^\dagger\otimes B^\dagger)\right]
\] 
is a stationary state for $\e\otimes \f$, where $K$ is a normalization factor. Note that this state is entangled by the positive partial transpose criteria \cite{mintert2009basic}, thus (2) holds.

Suppose $|\lambda(\e\otimes \f)|_{\infty}=1$. Then there are finitely many mutually orthogonal minimal stationary states $\{\xi_i\}_{i=1}^{m}$ for $\e\otimes \f$. By  \cite[Corollary 6.5]{wolf2012quantum}, it contradicts the fact that $A\otimes B$, $\rho_1\otimes \sigma_1$ and $\rho_2\otimes \sigma_2$ can all be linearly represented by $\{\xi_i\}_{i=1}^{m}$. This proves (1).
%
%
%
%
\hfill $\Box$

\begin{corollary}\label{cor:main}
    Given a quantum memory cell $\e$, $|\lambda(\e)|_{\infty}$ can be super-activated if and only if the multiset $\bar{\eta}(\e)$ of external eigenvalues is not empty. Furthermore,  $|\lambda(\e^{\otimes t})|_{\infty}$ is increasing with $t$.  
\end{corollary}
{\it Proof.} It suffices to note that the set of eigenvalues of $\e$ is closed under complex conjugate. Furthermore, if $n>m$, the noiseless subsystems of $\e^{\otimes m}$ is also the noiseless subsystems of $\e^{\otimes n}$, so 
$|\lambda(\e^{\otimes n})|_{\infty}\geq |\lambda(\e^{\otimes m})|_{\infty}$. 
\hfill $\Box$


Note that computing $\bar{\eta}(\e)$ is an easy linear algebra exercise. Thus super-activation of a given quantum memory cell $\e$ can be checked easily without finding an entangled stationary state or the noiseless subsystems of $\e^{\otimes t}$. 

Theorem \ref{Theo_ns_super} and Corollary \ref{cor:main} have some interesting implications. First, if we want to super-activate $|\lambda(\e)|_{\infty}$ by collective use  of $\e$, $\e$ must have at least two (mutually orthogonal) stationary quantum states; i.e. $|\lambda(\e)|>1$. This means that classical information can be stored in the memory cell $\e$. Therefore, such super-activation implies the preservation of classical information. This is in sharp contrast to the super-activation in zero-error communication over quantum channels: there exist quantum channels $\f_1$ and $\f_2$ such that both of them have vanishing zero-error classical capacity (meaning that classical information cannot be sent without errors), but the {zero-error quantum capacity} of $\f_1\otimes \f_2$ is positive (meaning that we can use it to transmit quantum information perfectly)~\cite{cubitt2012extreme}.
Secondly, if an entangled stationary state is found, then there is at least one useful noiseless subsystem in the whole memory cell system that can be used to store (entangled) quantum states. So, entanglement can be served as a signal for protecting quantum information, like the period in the super-activation of $|\lambda(\cdot)|$.  Thirdly, the quantities $|\lambda(\cdot)|_{\infty}$ and $|\lambda(\cdot)|$ are not multiplicable, i.e. in general $|\lambda(\e\otimes \f)|_{\infty}\not =|\lambda(\e)|_{\infty}\cdot |\lambda(\f)|_{\infty}$ and $|\lambda(\e\otimes \f)|\not =|\lambda(\e)|\cdot |\lambda(\f)|$.  This implies that the amount of information that can be preserved  through a quantum memory cell depends on what other memory cells are also available.

Mathematically, given a quantum memory cell $\e$, the shape is fully determined by its magnitude-one eigenvalues and the corresponding eigenvectors, and the super-activation of $|\lambda(\cdot)|$ and $|\lambda(\e)|_{\infty}$ is determined  by internal and external eigenvalues of $\e$, respectively. This indicates that eigenvalues with magnitude one have different roles in information storage.

Finally, we present a  simple example to show that the growth of the super-activation of $|\lambda(\cdot)|_{\infty}$ can be exponentially fast, and the speed of growth is independent on external eigenvalues. Therefore, the collective use of purely classical memory cells is an efficient method to preserve quantum information.

\begin{example}Let $\theta_1,\theta_2$ be real  numbers and $0<\theta_1\leq\theta_2< 2\pi$. We consider two quantum memory cells $\e_{k}(\rho)=(|0\rangle\langle 0 |+e^{i\theta_k}|1\rangle\langle 1 |)\rho(|0\rangle\langle 0 |+e^{-i\theta_{k}}|1\rangle\langle 1 |)+|2\rangle\langle 2 |\rho|2\rangle\langle 2 |$ on $\h=\textrm{lin.span}\{|0\rangle, |1\rangle,|2\rangle\}$ for $k=1,2$. Note that for each $k$, $$\e_{k}(\rho)=(|0\rangle\langle 0 |+e^{i\theta_k}|1\rangle\langle 1 |)(P_{0}+P_{1})\rho (P_{0}+P_{1})(|0\rangle\langle 0 |+e^{-i\theta_k}|1\rangle\langle 1 |)+|2\rangle\langle 2 |P_{2}\rho P_{2}|2\rangle\langle 2 |,$$ where $P_{j}$ is the projection onto lin.span$\{|j\rangle\}$ for $j\in\{0,1,2\}.$ So, we can restrict $\e_k$ onto lin.span$\{|0\rangle,|1\rangle\}$ when we only consider $|\lambda(\cdot)|_{\infty}$. Then the evolution is fully represented by  unitary matrices $\{U_k=diag(1,e^{i\theta_k})\}_{k=1,2}$. It is easy to compute $\lambda(\e_k)=(1,1,1)$, and $\bar{\eta}(\e_k)=\{e^{i\theta_k},e^{-i\theta_k}\}$ for $k=1,2$. By Theorem~\ref{Theo_ns_super}, $|\lambda(\e_1\otimes\e_2)|_{\infty}>1$ if and only if $\theta_1=\theta_2$ or $\theta_1=2\pi-\theta_2$. Now, we show that the growth speed of the super-activation is independent on $\theta_1, \theta_2$ and exponential with the number of collectively used memory cells in both cases. 
\begin{itemize}
    \item[(1)] $\theta_1=\theta_2$. Then let $U=U_{1}=U_2$ and $\e=\e_1=\e_2$. For any strictly positive integer $t$, $$U^{\otimes t}\simeq diag[I_{\binom{t}{0}},I_{\binom{t}{1}}e^{i\theta},\cdots,I_{\binom{t}{t}}e^{i t\theta} ]$$ where $I_{k}$ is the identity matrix with dimension $k$. Then $|\lambda(\e^{\otimes t})|_{\infty}>\binom{t}{t/2}$ if $t$ is even; otherwise, $|\lambda(\e^{\otimes t})|_{\infty}>\binom{t}{(t-1)/2}$.      
    \item[(2)] $\theta_{1}=2\pi-\theta_2$, i.e. $U_1=U_2^\dagger$. 
    For any strictly positive integer $t$, $|\lambda[(\e_1\otimes \e_2)^{\otimes t}]|_{\infty}\geq\binom{t}{t/2}^2$ if $t$ is even; otherwise $|\lambda[(\e_1\otimes \e_2)^{\otimes t}]|_{\infty}\geq\binom{t}{(t-1)/2}^2$.
\end{itemize}
Therefore, in the above cases, the growth speed is independent on $\theta_1,\theta_2$. Specifically, by Stirling's approximation, $\binom{t}{t/2}$ and $\binom{t}{(t-1)/2}$ both are growing up exponentially with $t$ and $|\lambda(\e^{\otimes t})|_{\infty}\leq n^t$ with dim$(\h)=n$, so 
    $|\lambda(\cdot)|_{\infty}$ has an exponential growth.
\end{example}

\section{Conclusion}
We proved that the existence of entangled stationary states of a given purely classical memory cell is necessary and sufficient for super-activating it to store quantum information, whereas a null memory cell can only be super-activated to preserve classical information.  We also proposed a simple method to check whether such super-activation happens by computing its external and internal eigenvalues, respectively. Moreover, once activated, the preserved quantum information may have exponential growth with the number of the used memory cells. This provides an efficient way to perfectly preserve quantum information even when the quantum memory cell is fully classical. 

At this moment, we only have a simple example showing the exponential growth of super-activation of the amount $|\lambda(\cdot)|_{\infty}$ of stored quantum information. In future research, we expect to give a general characterization of the growth speed of $|\lambda(\cdot)|_{\infty}$. 

A recent work \cite{jaques2018spectral} indicated that if we consider unital quantum memory cell $\e$ $(\e(I)=I)$ and use the decoherence-free subsystems storing quantum information, then the super-activation of quantum information cannot exist. So checking the existence for general quantum memory cells is also an interesting problem. 
\section*{Acknowledgements}
This work was partly supported by the National Key R$\&$D Program of China 
(Grant No: 2018YFA0306701), the National Natural Science Foundation of China  
(Grant No: 61832015) and the Australian Research Council (ARC) under grant Nos. DP160101652 and DP180100691.
\section*{References}
\bibliographystyle{plain}
\bibliography{QIC}

\section*{Appendix: Proof of Lemma~\ref{lem_eigen}}
\noindent
\begin{lemma}\label{lem_range_eigenvector}
    Let $\e$ be a quantum operation on $\h$ with the fixed-point decomposition $\h=\bigoplus_{k=1}^n(\h_{A_k}\otimes \h_{B_k})\oplus \mathcal{K}$, and $X$ an eigenvector corresponding to some $a\in\eta(\e)$. Then $X\in B(\mathcal{K}^\perp)$.
\end{lemma}
{\it Proof.} Following \cite[Proposition 6.3]{wolf2012quantum}, $\e_{\phi}$ is generated by $\e$ as follow.
Let $\{E_i\}$ be the  Kraus operators of $\e$, i.e. $\e(\cdot)=\sum_i E_i\cdot E_i^\dagger$. Then its matrix representation is defined to be $M=\sum_{i}E_{i}\otimes E_{i}^{*}$. Assume that $M=SJS^{-1}$ is the Jordan decomposition of $M$, where
\begin{eqnarray*}
J=\sum_{k=1}^{K}\lambda_{k}P_k+N_k,
\end{eqnarray*}
$N_{k}^{d_k}=0$ for some $d_k>0$, $N_{k}P_{k}=P_{k}N_{k}=N_{k}, P_{k}P_{l}=\delta_{kl}P_{k}, \textrm{tr}(P_{k})=d_k$, and  $\sum_{k}P_{k}=I$. 
Let $J_{\phi}:=\sum_{k:|\lambda_{k}|=1}P_{k}.$  Then we write $\e_\phi$ for the super-operator with the matrix representation $SJ_{\phi} S^{-1}$. 

By the definition, we first observe that $\e_{\phi}(X)=X$ for all $X$ with $\e(X)=aX$ for some $a\in \eta(\e)$. Then following \cite{ying2013reachability}, $\mathcal{K}$ is transitive, i.e. for all $\rho\in D(\h)$,
$\lim_{n\rightarrow \infty}\textrm{tr}(P\e^{n}(\rho))=0,$ 
where $P$ is the projection onto $\mathcal{K}$. Furthermore, \cite[Proposition 6.3]{wolf2012quantum} asserts that there exists an increasing sequence of integers $n_i$ such that $\e_\phi = \lim_{i\rightarrow \infty} \e^{n_i}$. Therefore, $\textrm{tr}(P\e_{\phi}(\rho))=0$ for all $\rho\in D(\h)$. The lemma follows by noting that $fix(\e_{\phi})$ has a decomposition similar to Eq.$(1)$ and $X$ can be linearly represented by a set of quantum states.
\hfill $\Box$

If we only want to compute eigenvalues of $\e$ with magnitude one and corresponding eigenvectors, then the restriction of $\e$ onto $\k^\perp$ is enough to be used by Lemma \ref{lem_range_eigenvector}. Furthermore, the following lemma gives a characterization of these eigenvalues when $\lambda(\e)=(1)$.
\begin{lemma}\label{lem_eigen_irreducible}
    For any quantum operation $\e$ with $\lambda(\e)=(1)$, we have $\eta(\e)=\{\exp(2\pi ik/p)\}_{k=0}^{p-1}$ for some integer $p$, and all elements in $\eta(\e)$  are internal.
\end{lemma}
{\it Proof.}
As $\lambda(\e)=(1)$, there is only one stationary state $\rho^*$ of $\e$. 
By restricting $\e$ onto $X=\textrm{supp}(\rho^*)$, $\e_{X}$ is \emph{irreducible}; that is, its shape is $(1)$ and its stationary state is of full-rank. Then with \cite[Theorem 6.6]{wolf2012quantum}, $\eta(\e)\supseteq \eta(\e_X)=\{\exp(2\pi ik/p)\}_{k=0}^{p-1}$ for some integer $p$, and the multiplicity of any eigenvalue in $\eta(\e_X)$ is 1. Following Lemma \ref{lem_range_eigenvector}, there are no other eigenvalues in $\eta(\e)$, so  $\eta(\e)= \eta(\e_X).$ 
\hfill $\Box$
\begin{corollary}
    Given a quantum operation $\e$, if $\lambda(\e)=(d)$ for some positive integer $d$, then  $\eta(\e)=\{\exp(2\pi ik/p)\}_{k=0}^{p-1}$ for some integer $p$,  and all elements in $\eta(\e)$ have multiplicity $d$. 
\end{corollary}

Recall that $p$ is the period of $\e$, i.e., $p=p(\e)$. 
\begin{lemma}\label{lem_periodic_state}
    For a quantum operation $\e$ with $\lambda(\e)=(1)$, there exists a set of mutually orthogonal quantum states $\{\rho_{i}\}_{i=0}^{p(\e)-1}$ such that $\e(\rho_{i})=\rho_{i\boxplus 1}$, where  $\boxplus$ denotes subtraction modulo $p(\e)$.
\end{lemma}
{\it Proof.}  Without loss of generality, we assume that $\e$ is irreducible. Otherwise, we restrict $\e$ onto the support of the stationary state. 
Let $d=p(\e)$. By \cite[Theorem 4]{guan2018decomposition},
there exists a set of mutually orthogonal subspaces $\{B_{i}\}_{i=0}^{d-1}$ such that $\h=\bigoplus_i B_{i}$ and for each $i$, $B_i$ is invariant under $\e^{d}$. Then for each $i$, $\e^{d}|_{B_{i}}$, the restriction of $\e^{d}$ onto $B_i$, has a limit state, i.e. there is a quantum state $\sigma_{i}\in D(B_{i})$ such that for all $\rho\in D(B_{i})$,
    $
    \lim_{n\rightarrow \infty}\e^{dn}|_{B_{i}}(\rho)=\sigma_{i}.
    $
    As $B_{i}$ is an invariant subspace of $\e^{d}$, this means
    $
    \lim_{n\rightarrow \infty}\e^{dn}(\rho)=\sigma_{i}.
    $
    From  \cite[Theorem 4]{guan2018decomposition}, we have $\e(\rho)\in D(B_{i\boxplus 1})$. So,  
    \begin{eqnarray*}
    \lim_{n\rightarrow \infty}\e^{dn}(\rho)=\sigma_{i} 
    \Rightarrow&&\lim_{n\rightarrow \infty}\e^{dn+1}(\rho)=\e(\sigma_{i}) \\
    \Rightarrow&&\lim_{n\rightarrow \infty}\e^{dn}(\e(\rho))=\e(\sigma_{i}) \\
    \Rightarrow&&\lim_{n\rightarrow \infty}\e^{dn}|_{B_{i\boxplus1}}(\e(\rho))=\e(\sigma_{i})\\
    \Rightarrow&&\sigma_{i\boxplus1}=\e(\sigma_{i}).
    \end{eqnarray*}\hfill $\Box$
\end{document}